\documentclass[a4paper,fleqn,usenatbib]{mnras}

\usepackage{mathptmx}

\usepackage[T1]{fontenc}
\usepackage{ae,aecompl}

\usepackage{graphicx}	
\usepackage{amsmath}	
\usepackage{amssymb}	

\newcommand{\pcm}{cm$^{-2}$}
\newcommand{\xmm}{\textit{XMM-Newton}}

\newcommand{\delcstat}{$\Delta$C-stat}

\title[Winds in 1H0707-495]{A stratified ultrafast outflow in 1H0707-495?}

\author[P Kosec et al.]{P. Kosec$^{1}$\thanks{E-mail: pk394@cam.ac.uk}, 
D. J. K. Buisson$^{1}$, 
M. L. Parker$^{2}$, 
C. Pinto$^{1}$, 
A. C. Fabian$^{1}$ and 
\newauthor D. J. Walton$^{1}$
\\
$^{1}$Institute of Astronomy, Madingley Road, CB3 0HA Cambridge, UK \\
$^{2}$European Space Agency (ESA), European Space Astronomy Centre (ESAC), E-28691 Villanueva de la Ca\~nada, Madrid, Spain
}

\date{Accepted 2018 August 24. Received 2018 August 22; in original form 2018 July 13}

\pubyear{2018}

\begin{document}
\label{firstpage}
\pagerange{\pageref{firstpage}--\pageref{lastpage}}
\maketitle

\begin{abstract}

Ultrafast outflows (UFOs) have recently been found in the spectra of a number of active galactic nuclei (AGN) and are strong candidates for driving AGN feedback. 1H0707-495 is a highly accreting narrow line Seyfert 1 and the second most X-ray variable bright AGN. Previous studies found evidence of blueshifted absorption at 0.1-0.2c in its spectrum. We perform a flux-resolved analysis of the full \xmm\ dataset on this AGN using both CCD and grating data, focusing on the low flux spectrum. We find strong evidence for an ultrafast outflow in absorption at $\sim$0.13c, with an ionisation parameter $\log(\xi$/erg~cm~s$^{-1})=4.3$. Surprisingly, we also detect blueshifted photoionised emission, with the velocity increasing at higher ionisation states, consistent with a trend that has been observed in the UV spectrum of this object. The bulk of the X-ray emitting material is moving at a velocity of 8000 km/s, with ionisation parameter $\log(\xi$/erg~cm~s$^{-1})=2.4$. The wind kinetic power inferred from the UFO absorption is comparable to that of the UV and X-ray emission features despite their different velocities and ionisation states, suggesting that we are viewing an energy-conserving wind slowing down and cooling at larger distances from the AGN.

\end{abstract}




\section{Introduction}

Ultrafast outflows (UFOs) are thought to be powerful winds from active galactic nuclei (AGN) with outflow velocities greater than 10000 km/s. They are most commonly identified through high-energy absorption features from \ion{Fe} {XXV/XXVI} in the 7-10 keV energy band \citep[e.g.][]{Tombesi+10}, consistent with highly ionized gas blueshifted at significant fractions of the speed of light. These features are generally thought to originate in winds magnetically or radiatively driven off the AGN accretion disc at high Eddington rates \citep{Pounds+03, Reeves+03, Fukumura+15}. If this scenario is correct, these winds are of great interest because they are very strong candidates for driving AGN feedback, as they couple with galactic gas much more efficiently than jets.

Narrow line Seyfert 1s (NLS1s) are a subclass of highly variable AGN that usually host lower-mass supermassive black holes ($10^{6}-10^{7} M_{\odot}$). They are thought to accrete at high Eddington fractions \citep[possibly super-Eddington,][]{Jin+12}. IRAS 13224-3809 (hereafter IRAS 13224) and 1H0707-495 (hereafter 1H0707) are the most X-ray variable objects from this class, varying by factors of tens within hours. These two objects share many similarities, including their broadband X-ray spectra, timing behaviour \citep{Zoghbi+10, Kara+13} and UV emission features \citep{Leighly+04a}.

\citet{Parker+16, Parker+17} recently found a variable $\sim$0.2c UFO in the spectrum of IRAS 13224. It is observed most clearly in the 7-10 keV band, but is also visible in the 0.4-2 keV Reflection Grating Spectrometer (RGS) data \citep{Pinto+18}. The UFO absorption is strongest when the flux of the AGN is low, and almost disappears in high flux states, varying on time-scales of ks. Given all the similarities between IRAS 13224 and 1H0707, it is reasonable to expect 1H0707 to also possess powerful winds.

Early studies of 1H0707 with \xmm\ already noticed a sharp drop in flux at 7 keV, which was interpreted as the blue wing of relativistically broadened iron K emission, or alternatively as blueshifted ionised iron absorption or the blue wing of a P Cygni wind profile \citep{Boller+02, Fabian+02, Done+07}. With larger datasets, it was possible to show that a major component of the continuum emission is indeed iron K and L reflection \citep{Fabian+09, Zoghbi+10}. \citet{Dauser+12} found absorption features, on top of smeared reflection, consistent with the Si, S, Ar and Ca Ly$\alpha$ lines in the long 2008 and 2010 XMM-Newton observing campaigns, with outflow velocities of 0.11c and 0.18c, respectively. \citet{Blustin+09} found double-peaked emission lines consistent with a broad-line region origin, but no signs of a UFO in the high-spectral resolution RGS data using the full 2008 campaign dataset. More recently, \citet{Hagino+16} re-analysed the archival \xmm\ data, finding clear evidence of an absorption feature at 8 keV, likely from blueshifted ionised iron.

Here, we perform a flux-resolved analysis of the full \xmm\ dataset including high spectral-resolution RGS data aiming to confirm the presence of absorption features from a UFO. We report the detection of blueshifted photoionized gas in emission with velocity increasing with ionisation state (in agreement with UV data) in the low flux 1H0707-495 spectrum. We also show strong evidence for the presence of an ultrafast outflow at 0.14c in the same spectrum. This work will be followed by a more technical paper describing the full dataset in detail showing the full flux and time resolved dataset.

\section{Observations and Data Reduction}

There is a wealth of data from multiple campaigns on this source. Here we use the full \xmm\ dataset \citep{Jansen+01}, spanning a time period from 2000 to 2010, excluding only observations shorter than 10 ks.

The data were reduced using a standard pipeline with SAS v16, CalDB as of April 2018. We use data from pn \citep{Struder+01} and RGS \citep{denHerder+01} instruments only. The high-background periods are filtered with a threshold of 0.4 counts/sec (in the 10 to 12 keV band) for pn and 0.15 counts/sec for RGS data. The source regions for pn were circles centred on the object with a radius of 20 arcsec (to maximise the signal-to-noise ratio in the 7-10 keV band), and the background regions were circles with a radius of 60 arcsec on the same chip as the source while avoiding the copper ring. RGS data was reduced with a standard pipeline (rgsproc task).

All spectra were converted into SPEX format for analysis in the SPEX fitting package \citep{Kaastra+96}. The pn data were grouped to at least 25 counts per bin and also binned by at least a factor 3 using the SPECGROUP procedure. RGS data was binned by a factor of 3 directly in SPEX to oversample the spectral resolution by about a factor of 3. The RGS spectral range used was 7 \AA\ (1.77 keV) to 31 \AA\ (0.4 keV), limited by the background; pn data were used between 1 and 10 keV to provide some overlap with the RGS energy band where the RGS count rate is low (<12 \AA).

\begin{figure*}
	\includegraphics[width=\textwidth]{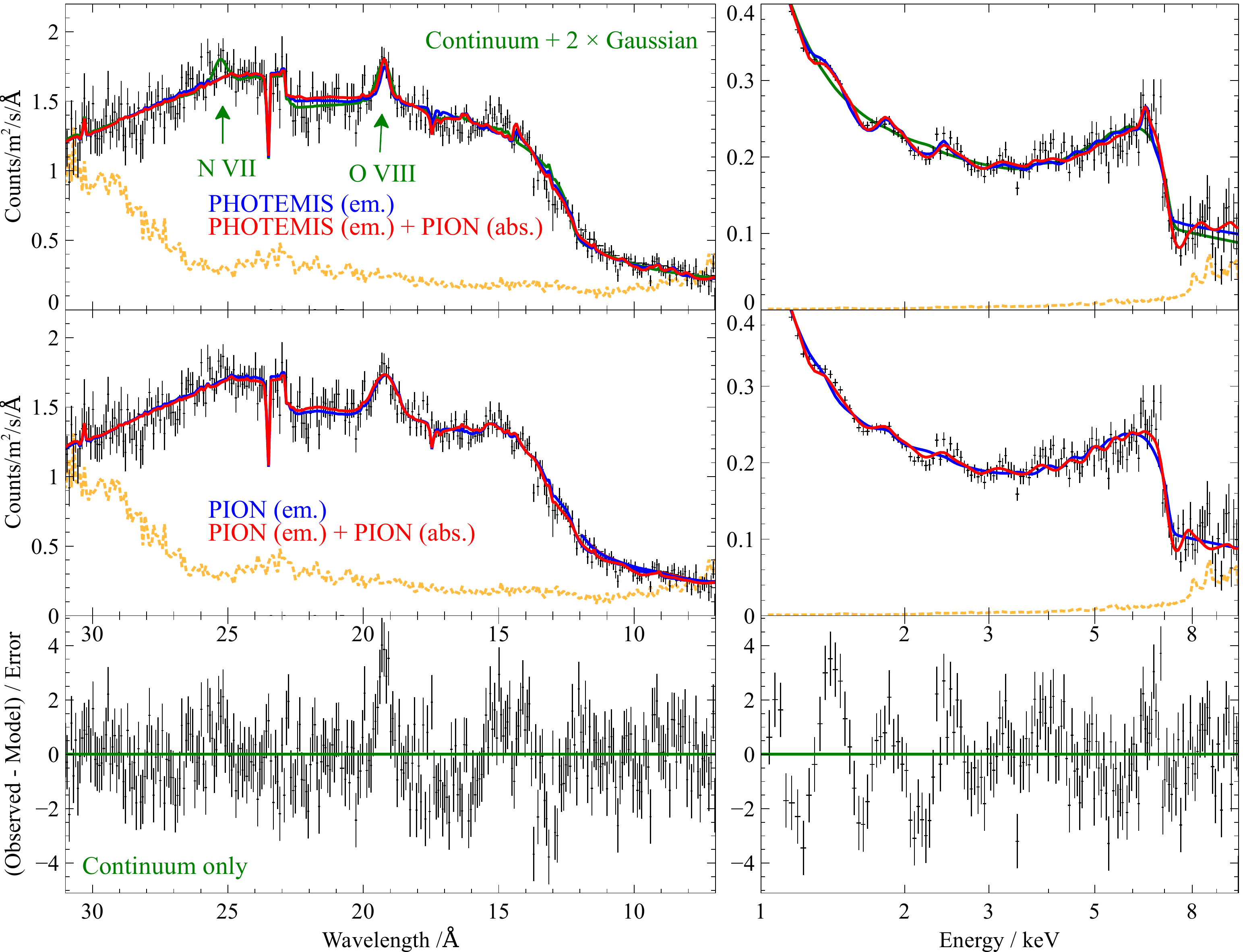}
    \caption{Low flux X-ray spectrum of 1H0707-495 using all \xmm\ data and both RGS (left subplots) and pn (right subplots) instruments. The top 2 rows of sub-plots contain the same data but fitted with different spectral models (Section \ref{results}). The top plots show a broadband continuum model plus 2 additional Gaussians (green), continuum plus \textsc{photemis} (blue), and continuum plus \textsc{photemis + pion} (red). The two middle plots contain fits with the continuum model plus \textsc{pion} (emission, blue), and \textsc{pion+pion} (both emission and absorption, in red). The bottom plots show the residuals to the broadband continuum fit without any blueshifted emission or absorption included. Note that the left plots are in Angstrom and the X-axis is linear, while the right plots are in keV with a logarithmic X-axis, both axes show the observed quantities (not rest-frame). The Y-axis shows flux, unfolded with the instrument response only, and is linear in all plots (but note the different units in left and right plots). Background is shown in orange colour. The RGS spectrum has been rebinned for plotting purposes.}
    \label{Spectrum}
\end{figure*}

We performed a flux-resolved reduction to analyse the source spectrum in the low flux state, which is the state in which the UFO absorption is strongest in IRAS 13224 \citep{Parker+16}. We extracted 3 spectra of the source in different flux states so that each has comparable statistics. This was done by taking a full lightcurve (bin size of 100 s) of the source during all the 15 \xmm\ observations and identifying the flux cut limits. Afterwards, the flux-resolved pn and RGS spectra were extracted for each observation, making sure that the extracted intervals for RGS and pn instruments are identical. Finally, all the individual observation spectra were stacked into 3 RGS and 3 pn spectra. The low-flux spectra are shown in Fig. \ref{Spectrum}. The total exposure (RGS1/RGS2) is around 500 ks, with a 0.3-10 keV average flux of $4.0 \times 10^{-12}$ erg~s$^{-1}$~cm$^{-2}$

We obtained the object redshift of $z=0.0405$ from the NASA/IPAC Extragalactic Database. The Galactic column density at the source position is around $N_{H}=4.6 \times 10^{20}$ \pcm\ \citep[Leiden/Argentine/Bonn Survey of Galactic HI,][]{Kalberla+05}. All the fits are performed in the SPEX fitting package using Cash statistics \citep{Cash+76}. All uncertainties are stated at the 1$\sigma$ level. Throughout the paper we assume solar abundances \citep{Lodders+09}.

\section{Methods and Results}
\label{results}

\subsection{Broadband Fitting}

The broadband (0.4-10 keV) spectrum was fitted with a phenomenological model composed of a powerlaw (\textsc{pow}), a soft ($\sim$0.1 keV) blackbody (\textsc{bb}) and 2 relativistic disc lines (\textsc{laor $\times$ gauss}), similar to the model used in \citet{Fabian+09}. The first Laor line describes smeared iron K reflection at 6 keV, the second one represents broadened iron L reflection around 1 keV and the blackbody models the soft excess of the source. The redshift of the source is represented with a \textsc{reds} model. All of the components are obscured by Galactic (mostly) neutral gas using the \textsc{hot} model in SPEX. An identical spectral model was used by \citet{Pinto+18} to describe the spectrum of IRAS 13224-3809 and results in an acceptable fit.

We describe the low flux spectrum of 1H0707-495 with this model, which results in a relatively poor fit (C-stat of 1554.60 for 935 D.o.F.), mostly because of a very strong emission feature around 19 \AA\ (0.65 keV), but also due to an emission residual around 25 \AA\ (0.5 keV) and multiple unresolved residuals in the pn spectrum. The best-fitting parameters are: powerlaw slope $\Gamma=2.61 \pm 0.02$, blackbody temperature $0.107 \pm 0.001$ keV, Laor line rest-frame energies $0.795_{-0.002}^{+0.007}$ keV and $5.64_{-0.03}^{+0.05}$ keV. We note that the best-fitting iron K line energy is impossibly low, which suggests probable absorption of its blue wing, as expected if a UFO is present. The inner disc radius given by the laor model is $1.29^{+0.04}_{-0.06}$, the emissivity of the disc is $7.3^{+0.4}_{-0.2}$ and its inclination $70.7^{+0.3}_{-0.2}$ degrees, in rough agreement with previous work. The fitted neutral column density is $7.3_{-0.5}^{+0.4} \times 10^{20}$ \pcm, significantly higher than Galactic \citep[in agreement with][]{Dauser+12}.

\subsection{Emission Lines Fitted with Gaussians}

From the fit statistics and the apparent residuals, it is evident that the simple broadband spectral model is insufficient. The strongest emission residuals are located at around 19 \AA\ (0.65 keV) and 25 \AA\ (0.50 keV) -- very close to the rest-frame positions of oxygen VIII and nitrogen VII transitions. There is no excess around 22 \AA\ (0.56 keV), where the oxygen VII transition is located. However, this part of the spectrum could be affected by neutral absorption in our and the host galaxy; also our exposure in the \ion{O}{VII} energy band is halved (missing chip in RGS2).

First, we fit these emission features with Gaussians (in addition to the broadband spectral model) to determine whether they are indeed rest-frame to the AGN host galaxy. We choose to couple the velocity widths of both lines, otherwise the \ion{N}{VII} line runs away to unreasonably large linewidths. The best-fitting line centroids (in the host galaxy rest-frame) are: $18.49 \pm 0.04$ \AA\ compared to \ion{O}{VIII} rest-frame wavelength of 18.97 \AA\, and $24.30 \pm 0.08$ \AA\ compared to \ion{N}{VII} rest-frame wavelength 24.78 \AA. The velocity width is $4200^{+900}_{-800}$ km/s. The line positions are neither rest-frame to the host nor our Galaxy and the lines are broad, therefore they must come from blueshifted gas near the AGN. The velocity of the gas emitting \ion{O}{VIII} is $7600^{+500}_{-600}$ km/s, and that of \ion{N}{VII} is $5800 \pm 1000$ km/s. Both Gaussians are highly statistically significant, \ion{O}{VIII} improving the fit statistics by $\Delta$C-stat $\sim88$ and \ion{N}{VII} by $\Delta$C-stat $\sim23$.

We note that both these emission features were already described by \citet{Blustin+09} in the 2008 campaign dataset, who also suggested the presence of redshifted components (both components interpreted to originate in the accretion disc at $\sim$1600 R$_{G}$). Using the full low-flux dataset, we do not see evidence for any redshifted emission with flux comparable to the blueshifted features, and instead interpret the blueshifted emission as a photoionised wind.

\subsection{Physical Models of Photoionised Emission}

The different blueshifts of the emission lines and the lack of \ion{O}{VII} mean that it is not possible to fit them with a single spectral model with just one ionisation parameter. In the following part, we mainly focus on the stronger of the two features, the O VIII emission line. We fit the spectrum with two spectral models describing photoionised emission physically: \textsc{photemis} and \textsc{pion}. We have also attempted to fit the spectrum with collisional ionisation models (emission originating in shocks) but they generally provide a significantly worse description of the features.

First we use the \textsc{photemis} model, which describes recombination and collisional excitation emission from a photoionised slab of plasma. It is implemented in the XSPEC package, and can be exported into SPEX as a table. We achieve the best fit for the ionisation parameter of $\log\xi=2.3$ (erg~cm~s$^{-1}$) and the velocity width of 3000 km/s. Adding the photoionised emission component is highly statistically significant and improves the fit to C-stat=1354 for 933 D.o.F. (\delcstat$=200$ for 4 additional D.o.F.). We note that this fit not only describes the O VIII feature, but also reproduces residuals in pn spectrum between 2 and 5 keV very well. The systematic velocity of the outflowing photoionised emitter is $8400^{+400}_{-500}$ km/s.

Secondly, we use a native SPEX model called \textsc{pion}. \textsc{pion} is a photoionisation code that self-consistently calculates the ionisation balance directly from the continuum fit and can reproduce both emission and absorption features. The disadvantage in using the model is a much higher computational cost. In this section we use the model to reproduce the emission residuals only (by setting the covering fraction $F_{\textrm{COV}}=0$ and the opening angle $\Omega=1$). We achieve the best-fitting solution for an ionisation parameter of $\log\xi=2.35^{+0.05}_{-0.03}$ and a velocity width of $6300^{+700}_{-500}$ km/s, with a systematic velocity of $8700_{-700}^{+400}$ km/s with respect to the host rest-frame. The column density of emitting gas is $3.5 \pm 0.3 \times 10^{21}$ \pcm\ and the fit improvement is \delcstat~$=164$ for 4 additional D.o.F.

\subsection{Ultrafast Outflow Absorption Signatures}
\label{outflow_abs}

The very low energy of the iron K line (broadened with a \textsc{laor} shape) suggests that its blue wing might be obscured - potentially by an ultrafast outflow. To test for this possibility, we add another component to our continuum + photoionised emission fit. We use a second \textsc{pion} component to describe the potential blueshifted absorption only (setting $F_{\textrm{COV}}=1$ and $\Omega=0$ to just model absorption features). Both absorption and emission cannot be described by a single \textsc{pion} component because the velocities of emitting and absorbing gas are likely different (absorption is usually much faster, $\sim$0.1c). The blueshifted emission is in this case modelled by \textsc{pion} ($F_{\textrm{COV}}=0$ and $\Omega=1$ as in the previous section). 

Adding the blueshifted absorber results in a significant fit improvement of \delcstat$=117$ for 4 additional degrees of freedom (column density, ionisation parameter, systematic and turbulent velocity), with $\textrm{C-stat}=1273$ for 927 D.o.F. The result is shown in Fig. \ref{Spectrum}. The best-fitting outflow velocity (relativistically-corrected) is $38400^{+700}_{-1000}$ km/s (0.13c) and the ionisation parameter $\log\xi=4.32^{+0.03}_{-0.02}$. The column density of the absorber is $2.4 \pm 0.2 \times 10^{23}$ \pcm\ and its turbulent velocity is $9600^{+900}_{-800}$ km/s. The parameters of the slow emitter are similar to those in the previous section: column density of $3.9_{-0.3}^{+0.5} \times 10^{21}$ \pcm, ionisation parameter of $\log\xi=2.44 \pm 0.03$, turbulent velocity of $6100_{-700}^{+600}$ km/s and systematic velocity of $8400^{+400}_{-700}$ km/s. We note that the energy of the iron K line is now $6.61 \pm 0.05$ keV - which seems plausible if the line is produced by a mix of neutral (6.4 keV) and ionised (6.7 keV) iron.

We verified that using a \textsc{photemis} component to describe the emission lines produces comparable results. The best-fitting solution is found at a speed of $\sim$0.14c and gives the fit statistics of 1268 for 929 d.o.f., a \delcstat$=86$ improvement over the model with \textsc{photemis} emission only. The UFO absorption velocity is $41900 \pm 800$ km/s, column density $3.9 \pm 0.1 \times 10^{23}$ \pcm, ionisation parameter $\log\xi=4.16 \pm 0.02$ and turbulent velocity $12200^{+800}_{-700}$ km/s. The continuum and emission parameters remain almost unchanged except for the best-fitting emitter ionisation parameter, which is $\log \xi=2.1$.

\begin{table*}
	\centering
	\caption{The best-fitting parameters of models of photoionised emission and absorption (see Sect. \ref{outflow_abs}).}
	\label{Final_results}
	\begin{tabular}{cccccccccc}
		\hline
		Model&\multicolumn{4}{c}{Emission}&&\multicolumn{4}{c}{Absorption}\\
		& V$_{\textrm{syst}}$ & N$_{\textrm{H}}$ & $\log(\xi)$ & V$_{\textrm{turb}}$ &\ & V$_{\textrm{syst}}$ & N$_{\textrm{H}}$ & $\log(\xi)$ & V$_{\textrm{turb}}$ \\
		& km~s$^{-1}$ & \pcm & erg~cm~s$^{-1}$ & km~s$^{-1}$ && km~s$^{-1}$ & \pcm & erg~cm~s$^{-1}$ & km~s$^{-1}$ \\
		\textsc{pion+pion}&$8400^{+400}_{-700}$&$3.9_{-0.3}^{+0.5} \times 10^{21}$&$2.44 \pm 0.03$&$6100_{-700}^{+600}$&&$38400^{+700}_{-1000}$&$(2.4 \pm 0.2) \times 10^{23}$&$4.32^{+0.03}_{-0.02}$&$9600^{+900}_{-800}$\\[0.1cm]
		\textsc{photemis+pion}&$8100^{+300}_{-400}$&*&2.1&3000&&$41900 \pm 800$&$(3.9 \pm 0.1) \times 10^{23}$&$4.16 \pm 0.02$&$12200^{+800}_{-700}$\\
		\hline
	\end{tabular}
	\\
	* It is not possible to extract an accurate value of the gas column density with \textsc{photemis}.
\end{table*}

A large fraction of the statistical significance upon adding the UFO absorption comes from the \ion{Fe}{XXV/XXVI} absorption (Fig. \ref{Spectrum}). In addition, the absorption lines of \ion{Mg}{xii}, \ion{Si}{xiv} and \ion{S}{xvi} are clearly fitted between 1.5 and 3 keV. Smaller absorption features are located at $\sim10$ \AA\ (\ion{Ne}{X}) and $\sim17$ \AA\ (\ion{O}{VIII}). We note that the features between 2 and 3 keV can also be fitted with \ion{S}{XV} emission instead of absorption (\textsc{photemis} emission fit only), although the first option is more plausible \citep[the features are also seen in IRAS 13224 where no strong photoionised emission is present,][]{Parker+17}. Additionally, the \textsc{photemis} fit also adds an iron emission feature at $\sim$6.4 keV, likely a blend of \ion{Fe}{XIX-XXI} lines.

To place the UFO solution on a firmer footing, we quantify the statistical fit improvement for the strongest individual UFO absorption lines: \ion{Fe}{XXV/XXVI}, \ion{S}{XVI}, \ion{Si}{XIV}, \ion{Mg}{XII}, \ion{Ne}{X} and \ion{O}{VIII}. We take the original broadband continuum and add a Gaussian at the energy where an absorption line is expected for the best-fitting UFO velocity, fit for its parameters, recovering the fit improvement \delcstat, its blueshift and turbulent velocity. The results are shown in Table \ref{Gaussian_absorption}. We note that a fit improvement of more than \delcstat~$=25$ is very high for an individual line \citep[including the look-elsewhere effect,][]{Kosec+18a} but Monte Carlo simulations would have to be performed to quantify the exact statistical significance. There are 2 likely explanations for the 7.6 keV residual - \ion{Fe}{XXV} or \ion{Fe}{XXVI}. At the ionisation level of $\log \xi=4.0-4.3$, the feature will most likely be a blend of both \ion{Fe}{XXV} and \ion{Fe}{XXVI}, but the majority of absorption will still originate from \ion{Fe}{XXV}. In conclusion, most of the absorption residuals are highly significant and support a coherent picture of a UFO moving at a velocity of $\sim$40000 km/s. We note that the velocity width of the \ion{Mg}{XII} residual is significantly lower than the widths of the other residuals. It is possible that the residual is affected by nearby slow emission transitions. This could also be the case for the 2 lowest energy transitions, \ion{Ne}{X} and \ion{O}{VIII}, where the blueshifted emission features are strongest and the simple broadband continuum does a poor job of reproducing the spectrum. Specifically, upon accounting for the emission features with a physical model, the underlying continuum changes slightly which can affect the true absorption line position and its depth (and hence the \delcstat\ fit improvement). Most notably, the absorption residual associated with \ion{O}{VIII} is located between 2 strong emission lines at 19 \AA\ (\ion{O}{VIII}) and 15 \AA\ (\ion{Fe}{XVII}) in observed frame (see Fig. \ref{Spectrum}). Therefore the best-fitting energies of these low energy (<2 keV) absorption lines should be taken with caution.

\begin{table*}
	\centering
	\caption{The strongest absorption residuals fitted with a Gaussian. Column (1) shows the most likely identification of the elemental transition, column (2) the fit improvement in \delcstat\ upon fitting the feature with a Gaussian, (3) lists the rest-frame energy of the transition and (4) the best-fitting energy of the Gaussian in the AGN rest-frame. Column (5) shows the inferred blueshift of the line, and (6) the turbulent velocity derived from the full width half maximum. The first two transitions detail different interpretations for the same Fe K feature.}
	\label{Gaussian_absorption}
	\begin{tabular}{cccccc} 
		\hline
		Transition &\delcstat &Rest-frame energy &Best-fitting energy &Blueshift &Turbulent width \\
		& &keV &keV &km~s$^{-1}$ &km~s$^{-1}$\\
		(1)  & (2) & (3) & (4) & (5) &(6) \\[0.1cm]
		\ion{Fe}{XXV} &44.68 &6.70 &$7.63 \pm 0.03 $&$38900 \pm 1100$ &$12800^{+1000}_{-1100}$\\[0.1cm]
		\ion{Fe}{XXVI} &44.68 &6.96	&$7.63 \pm 0.03$ &$27600 \pm 1100$ &$12800^{+1000}_{-1100}$\\[0.1cm]
		\ion{S}{XVI} &6.76 &2.62 &$2.97 \pm 0.06$ &$38000 \pm 6000$ &$21000 \pm 11000$\\[0.1cm]
		\ion{Si}{XIV} &41.01 &2.01 &$2.23 \pm 0.02$ &$32000 \pm 3000$ &$25000^{+6000}_{-7000}$\\[0.1cm]
		\ion{Mg}{XII} &25.72 &1.472 &$1.702 \pm 0.005$ &$43200 \pm 800$ &$<2100$\\[0.1cm]
		\ion{Ne}{X} &62.58 &1.022 &$1.188_{-0.014}^{+0.019}$ &$45000_{-5000}^{+4000}$ &$27000_{-6000}^{+4000}$\\[0.1cm]
		\ion{O}{VIII} &49.99 &0.654 &$0.765 \pm 0.006$ &$47000 \pm 2000$ &$17700_{-1900}^{+2200}$\\
		\hline
	\end{tabular}
	\\
\end{table*}

\subsection{Time and Flux Variability of the Emission and Absorption Features}

A possible caveat of using 10 years worth of data to perform a flux-resolved analysis is potential time variability of the emission and absorption features. Such variability could for example wash out the features or broaden them. Indeed, \citet{Dauser+12} find evidence for a shift in the UFO absorption velocity when comparing the datasets from two long \xmm\ campaigns on the source, in 2008 and 2010. On the other hand, the emission features are expected to be less variable (if they originate at larger distances from the central engine). To check that this is the case, we extract low-flux spectra from the 2008 and 2010 campaigns separately using the same flux limits obtained in the full dataset reduction.

\begin{figure}
	\includegraphics[width=\columnwidth]{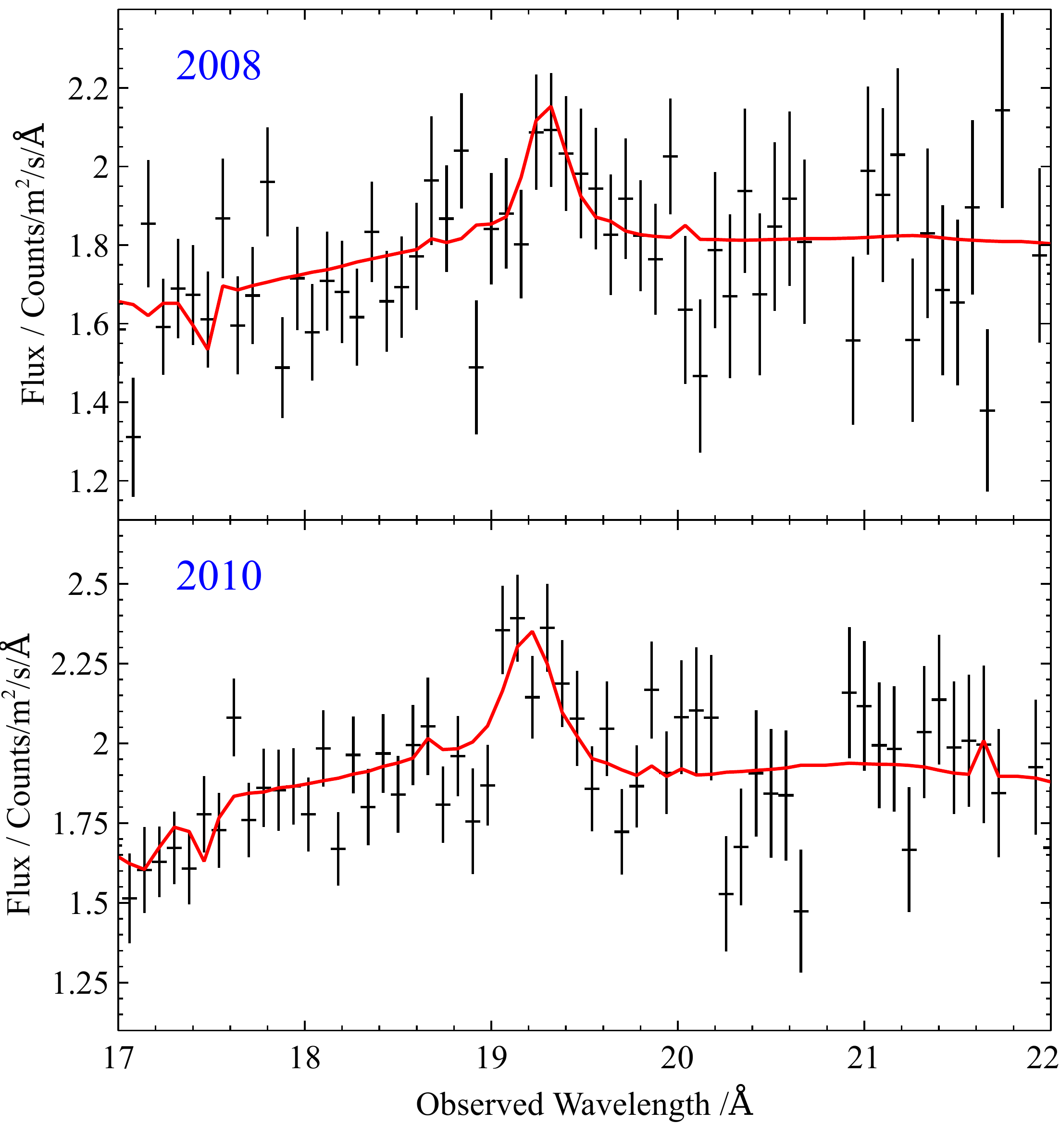}
    \caption{The \ion{O}{VIII} region of the 2008 (top) and 2010 (bottom) campaign low flux spectra extracted using the same flux limits used for the full dataset reduction. The X-axis shows the observed (not rest-frame) wavelength in \AA, and the Y-axis shows flux, unfolded with the instrument response only.}
    \label{Spectrum_2008_2010}
\end{figure}

We fit the 2 spectra with the photoionisation emission code \textsc{photemis+pion} in addition to the standard broadband continuum, which is computationally less expensive than using the \textsc{pion+pion} model. The best-fitting blueshifted emission parameters for the 2008 campaign data are: the systematic velocity of $7100 \pm 700$ km/s, $\log \xi=2.2$ and a turbulent velocity of 2000 km/s. The systematic velocity of the emitter during the 2010 campaign was $8500_{-300}^{+500}$ km/s, $\log \xi=2.0$ and a turbulent velocity of 2500 km/s. The strongest feature of the photoionised emission is \ion{O}{VIII} - the spectra of the \ion{O}{VIII} spectral band during 2008 and 2010 are shown in Fig. \ref{Spectrum_2008_2010} as a visual check.

We do therefore observe some variability between the two epochs, with approximately a 2$\sigma$ change in the outflow velocity. The ionisation parameter values need to be taken with caution as the data quality is much lower than using the full dataset. It does seem likely that co-adding the two epochs causes some broadening of the features. Nevertheless, the effects of stacking do not seem to be too severe and do not change the main conclusions of this analysis.

A full analysis of the medium and the high flux spectrum of 1H0707 is out of scope of this work. However, upon inspection of these spectra, it is evident that with increasing flux, the emission residuals decrease in relative strength compared to the broadband continuum. It is possible to fit them approximately with the same \textsc{photemis} model of photoionised emission used for the low flux spectrum (including its normalisation) which suggests that the emitting wind is independent of the AGN flux changes, as expected if it is located further from the AGN. The UFO absorption can be much more variable as seen in IRAS 13224 \citep{Parker+16}. In the case of 1H0707, with increasing flux the absorption residuals get weaker but are still present in the spectrum. Therefore it seems that the UFO does not disappear completely as might be the case for IRAS 13224.

\section{Discussion and Conclusions}

Our results show that using flux-resolved high-resolution X-ray spectroscopy to obtain a low-flux spectrum of the NLS1 1H0707-495 reveals blueshifted emission as well as strong evidence for an ultrafast outflow in absorption.

We find reasonable agreement between the different models used to describe the emission features. It is especially reassuring that the models agree on the velocity of about 8000 km/s and the ionisation parameter of $\log\xi\sim2.1-2.4$, from which the physical properties of the wind can be inferred. The turbulent velocity of gas is 3000-6000 km/s, and the column density is several times $10^{21}$ \pcm.

The UFO material is much more ionised with $\log\xi\sim4.3$, column density of $3 \times 10^{23}$ \pcm, turbulent velocity of 10000-12000 km/s and systematic speed of 39000-42000 km/s (0.13-0.14c). The parameters of the UFO do not stand out if compared to other AGN, where velocities between 0.1 and 0.3c are commonly observed \citep{Tombesi+10,Parker+18}. The ionisation parameter and column density are very similar to the UFO observed in IRAS 13224, although the velocity is smaller \citep[0.2-0.25c in IRAS 13224, see][]{Parker+17}. \citet{Dauser+12} analysed the 2010 and 2012 campaigns on 1H0707 separately and found evidence for a UFO velocity shift from 0.11c to 0.18c. Our result of 0.14c using the low flux state of the full dataset (2000 to 2011) therefore averages over this possible shift (however the full dataset is necessary to analyse the emission features). This is likely captured in the rather high velocity width of the UFO lines ($\sim$10000 km/s). A velocity shift would also decrease the UFO detection significance, which is nevertheless still very high with \delcstat$=85-115$ for 4 D.o.F. \citep[see][to see how Monte Carlo simulations can be used to determine the precise statistical significance]{Kosec+18a,Kosec+18b}. In each case, it is encouraging that including the absorber raises the rest-frame energy of the smeared iron K emission line from an unphysically low value of $\sim$5.6 keV to $\sim$6.6 keV, regardless of the absorption model used.

With this discovery, we are adding 1H0707 to a list of known AGN with strong evidence for multiphase outflows. Detection of two UFOs in absorption at 0.06c and 0.13c was reported using hard X-ray, soft X-ray and also UV data in the spectrum of another NLS1 PG 1211+143 \citep{Pounds+16a,Pounds+16b,Kriss+18}. Similarly, the spectrum of PDS 456 shows evidence for two very fast UFOs at 0.25c and at 0.46c \citep{Nardini+15,Reeves+18}. Other examples are IRAS 17020+4544, with evidence for up to 5 UFO phases \citep{Longinotti+15}, and IRAS F11119+3257, where X-ray UFO absorption as well as a kpc-scale molecular outflow (with comparable energetics) is observed \citep{Tombesi+15}. However, we note two important differences between these objects and 1H0707. First, all the aforementioned outflows were detected in absorption only. To our knowledge, 1H0707 is currently the only known object to show significantly blueshifted X-ray absorption and emission at the same time. NGC 4051 shows evidence for some blueshifted X-ray emission \citep{Pounds+11}, but much slower at $\sim$750 km/s. The second difference is in the outflow velocities - while most UFOs in absorption achieve sub-relativistic speeds of 0.05-0.5c, here the blueshifted emission is significantly slower at <10000 km/s.

It is useful to estimate the total kinetic power of the outflowing gas. Here we estimate the power separately for both absorbing and emitting gas and compare them, following the steps of \citet{Pinto+17} and \citet{Kosec+18a}. The wind power is $\dot{E}_{\textrm{kin}}=\frac{1}{2}\dot{M}u^{2}$ where $u$ is the wind velocity. The outflow rate $\dot{M}$ is then determined using the ionisation parameter $\xi$: $\dot{M}=4\pi\Omega C_{\textrm{V}} m_{\textrm{H}} \mu L_{\textrm{ion}} u/\xi$ where $\Omega$ is the solid angle of the outflow as a fraction of $4\pi$, $C_{\textrm{V}}$ is the volume filling factor defining how clumpy the wind is and $L_{\textrm{ion}}$ is the ionising luminosity. $m_{\textrm{H}}$ is the hydrogen (proton) mass and $\mu$ the mean atomic weight ($\sim$1.2 if the abundances are solar). The mechanical power of the outflow is therefore: $\dot{E}_{\textrm{kin}}=2\pi\Omega C_{\textrm{V}} L_{\textrm{ion}} m_{\textrm{H}} \mu u^{3}/\xi$.

For the ultrafast absorber, using the best-fitting values from the \textsc{pion} + \textsc{pion} fit, we obtain $\dot{E}_{\textrm{kin}}=~34^{+4}_{-5}~\Omega C_{\textrm{V}} L_{\textrm{ion}}$. Repeating the same calculation for the photoionised emitter (from the same spectral fit) gives $\dot{E}_{\textrm{kin}}=~27_{-7}^{+6}~\Omega C_{\textrm{V}} L_{\textrm{ion}}$. If the ionising luminosity, solid angle and filling factor of the outflows are comparable, these are two strikingly similar estimates despite completely different velocities and ionisation parameters. However, we note that the volume filling factor, C$_{\textrm{V}}$, of both gas phases is highly uncertain. If for instance the lower ionisation wind component is in form of compact clumps, its C$_{\textrm{V}}$ could be much smaller than the one of the UFO absorbing gas. Then the kinetic power of the soft X-ray emitting material could be significantly lower than estimated above.

\begin{figure}
	\includegraphics[width=\columnwidth]{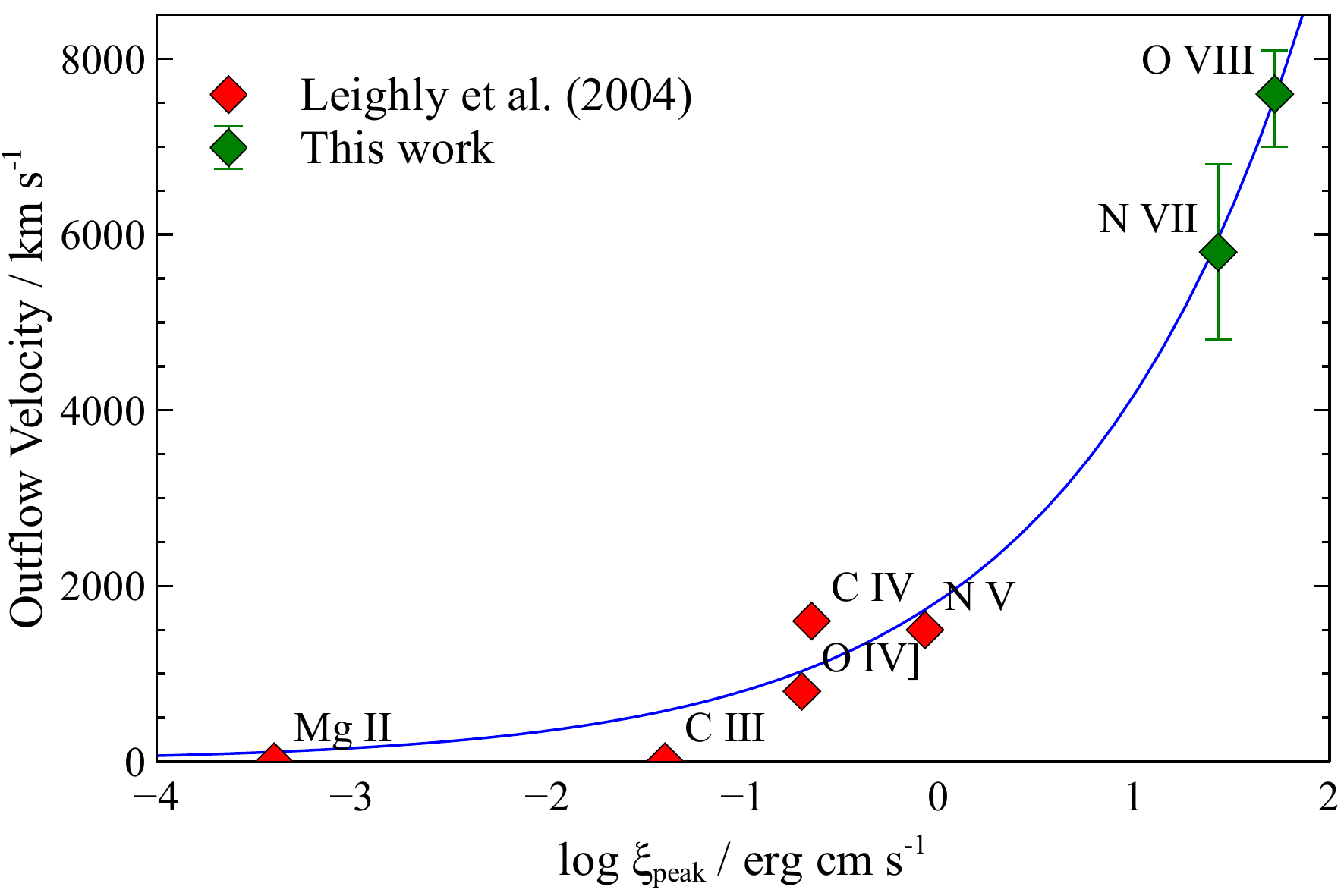}
    \caption{The outflow velocity of different ions versus the ionisation parameter at which the abundance of each partially ionised ion in a photoionised plasma peaks \citep[the SPEX value from][]{Mehdipour+16}. Values for the low-ionisation transitions were taken from \citet{Leighly+04a,Leighly+04b}. The blue curve is the best fit function to ions with non-zero velocity, in form $v=b\times(\xi$/erg~cm~s$^{-1})^{a}$, where $a=0.36 \pm 0.04$ and $b=(1800 \pm 300)$ km~s$^{-1}$.}
    \label{Ion_velocities}
\end{figure}

\citet{Leighly+04a} performed a UV spectral study of 1H0707 and IRAS 13224 and found that while low ionisation lines such as \ion{Mg}{II} and \ion{C}{III} appear to be rest-frame and of disc origin, higher ionisation lines \ion{Si}{IV}, \ion{O}{IV]} and \ion{C}{IV} are blueshifted at up to $\sim2000$ km/s. Here we have discovered an extension of the same trend in X-rays with 2 high ionisation transitions, \ion{N}{VII} and \ion{O}{VIII}, at much higher velocities (see Fig. \ref{Ion_velocities}). Curiously, no \ion{O}{VII} emission is seen, which could be due to imperfectly modelled absorption. If we fit the ions with non-zero velocity with a function in form $v=b\times\xi^{a}$, where $v$ is velocity, $\xi$ is the ionisation parameter and $a$ and $b$ are constants, the best-fitting powerlaw slope is $a=0.36 \pm 0.04$ \citep[we note that we choose arbitrary 500 km/s errorbars on UV ion velocities due to a lack of uncertainties in][]{Leighly+04a}. This suggests that $\frac{v^{3}}{\xi}$ is consistent with being constant and hence that energy is being conserved if the ion emission comes from the same wind which produces the UFO absorption and the soft X-ray emission.

In conclusion, we present strong evidence of ultrafast absorption as well as slower blueshifted emission in the X-ray spectrum of 1H0707. The trend of increasing velocities of higher ionised ions and the possible similar kinetic powers of UV, soft X-ray emitters and UFO absorbers suggest that we are witnessing the evolution of a stratified, kinetic energy-conserving wind. It would likely be launched close to the central accretor by radiation pressure, especially if the mass accretion rate is around or above the Eddington limit \citep[which is probably the case for 1H0707,][]{Jin+12}, and leave an imprint in form of UFO absorption. During the expansion, the wind would cool and slow down upon interaction with surrounding material, imprint the soft X-ray and UV spectra, and eventually deposit its kinetic energy at much larger (kpc) scales to produce AGN feedback. We note that this is not to be confused with the energy conserving feedback, which occurs on much larger (kpc) scales when the outflowing gas is no longer Compton-cooled by the AGN radiation and it expands adiabatically.

An alternative solution is that the blueshifted soft X-ray lines are the extension of an accelerating line-driven wind, proposed by \citep{Leighly+04b} based on the UV features only. The UFO absorption could then be completely unrelated, its kinetic power only coincidentally similar to that of the remaining features. It is also possible that the observed UFO absorption is not in fact signature of a real outflow, but rather blueshifted absorption of the relativistic reflection spectrum by an extended corotating atmosphere of small clouds above the disk (Fabian et al. 2018, submitted).

\section*{Acknowledgements}

We are grateful to the anonymous referee for useful comments that significantly improved the quality of the paper. PK and DJKB acknowledge support from the STFC. MLP is supported by a European Space Agency (ESA) research fellowship. CP and ACF acknowledge support from ERC Advanced Grant Feedback 340442. DJW acknowledges support from STFC Ernest Rutherford fellowships. This work is based on observations obtained with XMM-Newton, an ESA science mission funded by ESA Member States and USA (NASA). This research has made use of the NASA/IPAC Extragalactic Database (NED) which is operated by the Jet Propulsion Laboratory, California Institute of Technology, under contract with the National Aeronautics and Space Administration.





\bibliographystyle{mnras}
\bibliography{References} 

\begin{thebibliography}{}
\makeatletter
\relax
\def\mn@urlcharsother{\let\do\@makeother \do\$\do\&\do\#\do\^\do\_\do\%\do\~}
\def\mn@doi{\begingroup\mn@urlcharsother \@ifnextchar [ {\mn@doi@}
  {\mn@doi@[]}}
\def\mn@doi@[#1]#2{\def\@tempa{#1}\ifx\@tempa\@empty \href
  {http://dx.doi.org/#2} {doi:#2}\else \href {http://dx.doi.org/#2} {#1}\fi
  \endgroup}
\def\mn@eprint#1#2{\mn@eprint@#1:#2::\@nil}
\def\mn@eprint@arXiv#1{\href {http://arxiv.org/abs/#1} {{\tt arXiv:#1}}}
\def\mn@eprint@dblp#1{\href {http://dblp.uni-trier.de/rec/bibtex/#1.xml}
  {dblp:#1}}
\def\mn@eprint@#1:#2:#3:#4\@nil{\def\@tempa {#1}\def\@tempb {#2}\def\@tempc
  {#3}\ifx \@tempc \@empty \let \@tempc \@tempb \let \@tempb \@tempa \fi \ifx
  \@tempb \@empty \def\@tempb {arXiv}\fi \@ifundefined
  {mn@eprint@\@tempb}{\@tempb:\@tempc}{\expandafter \expandafter \csname
  mn@eprint@\@tempb\endcsname \expandafter{\@tempc}}}

\bibitem[\protect\citeauthoryear{{Blustin} \& {Fabian}}{{Blustin} \&
  {Fabian}}{2009}]{Blustin+09}
{Blustin} A.~J.,  {Fabian} A.~C.,  2009, \mn@doi [\mnras]
  {10.1111/j.1745-3933.2009.00746.x}, \href
  {http://adsabs.harvard.edu/abs/2009MNRAS.399L.169B} {399, L169}

\bibitem[\protect\citeauthoryear{{Boller} et~al.,}{{Boller}
  et~al.}{2002}]{Boller+02}
{Boller} T.,  et~al., 2002, \mn@doi [\mnras]
  {10.1046/j.1365-8711.2002.05040.x}, \href
  {http://adsabs.harvard.edu/abs/2002MNRAS.329L...1B} {329, L1}

\bibitem[\protect\citeauthoryear{{Cash}}{{Cash}}{1979}]{Cash+76}
{Cash} W.,  1979, \mn@doi [\apj] {10.1086/156922}, \href
  {http://adsabs.harvard.edu/abs/1979ApJ...228..939C} {228, 939}

\bibitem[\protect\citeauthoryear{{Dauser} et~al.,}{{Dauser}
  et~al.}{2012}]{Dauser+12}
{Dauser} T.,  et~al., 2012, \mn@doi [\mnras]
  {10.1111/j.1365-2966.2011.20356.x}, \href
  {http://adsabs.harvard.edu/abs/2012MNRAS.422.1914D} {422, 1914}

\bibitem[\protect\citeauthoryear{{Done}, {Sobolewska}, {Gierli{\'n}ski}  \&
  {Schurch}}{{Done} et~al.}{2007}]{Done+07}
{Done} C.,  {Sobolewska} M.~A.,  {Gierli{\'n}ski} M.,   {Schurch} N.~J.,  2007,
  \mn@doi [\mnras] {10.1111/j.1745-3933.2006.00255.x}, \href
  {http://adsabs.harvard.edu/abs/2007MNRAS.374L..15D} {374, L15}

\bibitem[\protect\citeauthoryear{{Fabian}, {Ballantyne}, {Merloni}, {Vaughan},
  {Iwasawa}  \& {Boller}}{{Fabian} et~al.}{2002}]{Fabian+02}
{Fabian} A.~C.,  {Ballantyne} D.~R.,  {Merloni} A.,  {Vaughan} S.,  {Iwasawa}
  K.,   {Boller} T.,  2002, \mn@doi [\mnras]
  {10.1046/j.1365-8711.2002.05419.x}, \href
  {http://adsabs.harvard.edu/abs/2002MNRAS.331L..35F} {331, L35}

\bibitem[\protect\citeauthoryear{{Fabian} et~al.,}{{Fabian}
  et~al.}{2009}]{Fabian+09}
{Fabian} A.~C.,  et~al., 2009, \mn@doi [\nat] {10.1038/nature08007}, \href
  {http://adsabs.harvard.edu/abs/2009Natur.459..540F} {459, 540}

\bibitem[\protect\citeauthoryear{{Fukumura}, {Tombesi}, {Kazanas}, {Shrader},
  {Behar}  \& {Contopoulos}}{{Fukumura} et~al.}{2015}]{Fukumura+15}
{Fukumura} K.,  {Tombesi} F.,  {Kazanas} D.,  {Shrader} C.,  {Behar} E.,
  {Contopoulos} I.,  2015, \mn@doi [\apj] {10.1088/0004-637X/805/1/17}, \href
  {http://adsabs.harvard.edu/abs/2015ApJ...805...17F} {805, 17}

\bibitem[\protect\citeauthoryear{{Hagino}, {Odaka}, {Done}, {Tomaru},
  {Watanabe}  \& {Takahashi}}{{Hagino} et~al.}{2016}]{Hagino+16}
{Hagino} K.,  {Odaka} H.,  {Done} C.,  {Tomaru} R.,  {Watanabe} S.,
  {Takahashi} T.,  2016, \mn@doi [\mnras] {10.1093/mnras/stw1579}, \href
  {http://adsabs.harvard.edu/abs/2016MNRAS.461.3954H} {461, 3954}

\bibitem[\protect\citeauthoryear{{Jansen} et~al.,}{{Jansen}
  et~al.}{2001}]{Jansen+01}
{Jansen} F.,  et~al., 2001, \mn@doi [\aap] {10.1051/0004-6361:20000036}, \href
  {http://adsabs.harvard.edu/abs/2001A%26A...365L...1J} {365, L1}

\bibitem[\protect\citeauthoryear{{Jin}, {Ward}, {Done}  \& {Gelbord}}{{Jin}
  et~al.}{2012}]{Jin+12}
{Jin} C.,  {Ward} M.,  {Done} C.,   {Gelbord} J.,  2012, \mn@doi [\mnras]
  {10.1111/j.1365-2966.2011.19805.x}, \href
  {http://adsabs.harvard.edu/abs/2012MNRAS.420.1825J} {420, 1825}

\bibitem[\protect\citeauthoryear{{Kaastra}, {Mewe}  \&
  {Nieuwenhuijzen}}{{Kaastra} et~al.}{1996}]{Kaastra+96}
{Kaastra} J.~S.,  {Mewe} R.,   {Nieuwenhuijzen} H.,  1996, in {Yamashita} K.,
  {Watanabe} T.,  eds, UV and X-ray Spectroscopy of Astrophysical and
  Laboratory Plasmas. pp 411--414

\bibitem[\protect\citeauthoryear{{Kalberla}, {Burton}, {Hartmann}, {Arnal},
  {Bajaja}, {Morras}  \& {P{\"o}ppel}}{{Kalberla} et~al.}{2005}]{Kalberla+05}
{Kalberla} P.~M.~W.,  {Burton} W.~B.,  {Hartmann} D.,  {Arnal} E.~M.,  {Bajaja}
  E.,  {Morras} R.,   {P{\"o}ppel} W.~G.~L.,  2005, \mn@doi [\aap]
  {10.1051/0004-6361:20041864}, \href
  {http://adsabs.harvard.edu/abs/2005A%26A...440..775K} {440, 775}

\bibitem[\protect\citeauthoryear{{Kara}, {Fabian}, {Cackett}, {Miniutti}  \&
  {Uttley}}{{Kara} et~al.}{2013}]{Kara+13}
{Kara} E.,  {Fabian} A.~C.,  {Cackett} E.~M.,  {Miniutti} G.,   {Uttley} P.,
  2013, \mn@doi [\mnras] {10.1093/mnras/stt024}, \href
  {http://adsabs.harvard.edu/abs/2013MNRAS.430.1408K} {430, 1408}

\bibitem[\protect\citeauthoryear{{Kosec}, {Pinto}, {Walton}, {Fabian},
  {Bachetti}, {F{\"u}rst}  \& {Grefenstette}}{{Kosec}
  et~al.}{2018a}]{Kosec+18b}
{Kosec} P.,  {Pinto} C.,  {Walton} D.~J.,  {Fabian} A.~C.,  {Bachetti} M.,
  {F{\"u}rst} F.,   {Grefenstette} B.~W.,  2018a, preprint, \href
  {http://adsabs.harvard.edu/abs/2018arXiv180302367K} {} (\mn@eprint {arXiv}
  {1803.02367})

\bibitem[\protect\citeauthoryear{{Kosec}, {Pinto}, {Fabian}  \&
  {Walton}}{{Kosec} et~al.}{2018b}]{Kosec+18a}
{Kosec} P.,  {Pinto} C.,  {Fabian} A.~C.,   {Walton} D.~J.,  2018b, \mn@doi
  [\mnras] {10.1093/mnras/stx2695}, \href
  {http://adsabs.harvard.edu/abs/2018MNRAS.473.5680K} {473, 5680}

\bibitem[\protect\citeauthoryear{{Kriss}, {Lee}, {Danehkar}, {Nowak}, {Fang},
  {Hardcastle}, {Neilsen}  \& {Young}}{{Kriss} et~al.}{2018}]{Kriss+18}
{Kriss} G.~A.,  {Lee} J.~C.,  {Danehkar} A.,  {Nowak} M.~A.,  {Fang} T.,
  {Hardcastle} M.~J.,  {Neilsen} J.,   {Young} A.,  2018, \mn@doi [\apj]
  {10.3847/1538-4357/aaa42b}, \href
  {http://adsabs.harvard.edu/abs/2018ApJ...853..166K} {853, 166}

\bibitem[\protect\citeauthoryear{{Leighly}}{{Leighly}}{2004}]{Leighly+04b}
{Leighly} K.~M.,  2004, \mn@doi [\apj] {10.1086/422089}, \href
  {http://adsabs.harvard.edu/abs/2004ApJ...611..125L} {611, 125}

\bibitem[\protect\citeauthoryear{{Leighly} \& {Moore}}{{Leighly} \&
  {Moore}}{2004}]{Leighly+04a}
{Leighly} K.~M.,  {Moore} J.~R.,  2004, \mn@doi [\apj] {10.1086/422088}, \href
  {http://adsabs.harvard.edu/abs/2004ApJ...611..107L} {611, 107}

\bibitem[\protect\citeauthoryear{{Lodders}, {Palme}  \& {Gail}}{{Lodders}
  et~al.}{2009}]{Lodders+09}
{Lodders} K.,  {Palme} H.,   {Gail} H.-P.,  2009, \mn@doi [Landolt
  B{\"o}rnstein] {10.1007/978-3-540-88055-4_34}, \href
  {http://adsabs.harvard.edu/abs/2009LanB...4B...44L} {}

\bibitem[\protect\citeauthoryear{{Longinotti}, {Krongold}, {Guainazzi},
  {Giroletti}, {Panessa}, {Costantini}, {Santos-Lleo}  \&
  {Rodriguez-Pascual}}{{Longinotti} et~al.}{2015}]{Longinotti+15}
{Longinotti} A.~L.,  {Krongold} Y.,  {Guainazzi} M.,  {Giroletti} M.,
  {Panessa} F.,  {Costantini} E.,  {Santos-Lleo} M.,   {Rodriguez-Pascual} P.,
  2015, \mn@doi [\apj] {10.1088/2041-8205/813/2/L39}, \href
  {https://ui.adsabs.harvard.edu/#abs/2015ApJ...813L..39L} {813, L39}

\bibitem[\protect\citeauthoryear{{Mehdipour}, {Kaastra}  \&
  {Kallman}}{{Mehdipour} et~al.}{2016}]{Mehdipour+16}
{Mehdipour} M.,  {Kaastra} J.~S.,   {Kallman} T.,  2016, \mn@doi [\aap]
  {10.1051/0004-6361/201628721}, \href
  {http://adsabs.harvard.edu/abs/2016A%26A...596A..65M} {596, A65}

\bibitem[\protect\citeauthoryear{{Nardini} et~al.,}{{Nardini}
  et~al.}{2015}]{Nardini+15}
{Nardini} E.,  et~al., 2015, \mn@doi [Science] {10.1126/science.1259202}, \href
  {http://adsabs.harvard.edu/abs/2015Sci...347..860N} {347, 860}

\bibitem[\protect\citeauthoryear{{Parker} et~al.,}{{Parker}
  et~al.}{2017a}]{Parker+17}
{Parker} M.~L.,  et~al., 2017a, \mn@doi [\mnras] {10.1093/mnras/stx945}, \href
  {http://adsabs.harvard.edu/abs/2017MNRAS.469.1553P} {469, 1553}

\bibitem[\protect\citeauthoryear{{Parker} et~al.,}{{Parker}
  et~al.}{2017b}]{Parker+16}
{Parker} M.~L.,  et~al., 2017b, \mn@doi [\nat] {10.1038/nature21385}, \href
  {http://adsabs.harvard.edu/abs/2017Natur.543...83P} {543, 83}

\bibitem[\protect\citeauthoryear{{Parker}, {Buisson}, {Jiang}, {Gallo}, {Kara},
  {Matzeu}  \& {Walton}}{{Parker} et~al.}{2018}]{Parker+18}
{Parker} M.~L.,  {Buisson} D.~J.~K.,  {Jiang} J.,  {Gallo} L.~C.,  {Kara} E.,
  {Matzeu} G.~A.,   {Walton} D.~J.,  2018, \mn@doi [\mnras]
  {10.1093/mnrasl/sly096}, \href
  {http://adsabs.harvard.edu/abs/2018MNRAS.tmpL..98P} {}

\bibitem[\protect\citeauthoryear{{Pinto} et~al.,}{{Pinto}
  et~al.}{2017}]{Pinto+17}
{Pinto} C.,  et~al., 2017, \mn@doi [\mnras] {10.1093/mnras/stx641}, \href
  {http://adsabs.harvard.edu/abs/2017MNRAS.468.2865P} {468, 2865}

\bibitem[\protect\citeauthoryear{{Pinto} et~al.,}{{Pinto}
  et~al.}{2018}]{Pinto+18}
{Pinto} C.,  et~al., 2018, \mn@doi [\mnras] {10.1093/mnras/sty231}, \href
  {http://adsabs.harvard.edu/abs/2018MNRAS.476.1021P} {476, 1021}

\bibitem[\protect\citeauthoryear{{Pounds} \& {Vaughan}}{{Pounds} \&
  {Vaughan}}{2011}]{Pounds+11}
{Pounds} K.~A.,  {Vaughan} S.,  2011, \mn@doi [\mnras]
  {10.1111/j.1365-2966.2011.18866.x}, \href
  {http://adsabs.harvard.edu/abs/2011MNRAS.415.2379P} {415, 2379}

\bibitem[\protect\citeauthoryear{{Pounds}, {Reeves}, {King}, {Page}, {O'Brien}
  \& {Turner}}{{Pounds} et~al.}{2003}]{Pounds+03}
{Pounds} K.~A.,  {Reeves} J.~N.,  {King} A.~R.,  {Page} K.~L.,  {O'Brien}
  P.~T.,   {Turner} M.~J.~L.,  2003, \mn@doi [\mnras]
  {10.1046/j.1365-8711.2003.07006.x}, \href
  {http://adsabs.harvard.edu/abs/2003MNRAS.345..705P} {345, 705}

\bibitem[\protect\citeauthoryear{{Pounds}, {Lobban}, {Reeves}  \&
  {Vaughan}}{{Pounds} et~al.}{2016a}]{Pounds+16a}
{Pounds} K.,  {Lobban} A.,  {Reeves} J.,   {Vaughan} S.,  2016a, \mn@doi
  [\mnras] {10.1093/mnras/stw165}, \href
  {http://adsabs.harvard.edu/abs/2016MNRAS.457.2951P} {457, 2951}

\bibitem[\protect\citeauthoryear{{Pounds}, {Lobban}, {Reeves}, {Vaughan}  \&
  {Costa}}{{Pounds} et~al.}{2016b}]{Pounds+16b}
{Pounds} K.~A.,  {Lobban} A.,  {Reeves} J.~N.,  {Vaughan} S.,   {Costa} M.,
  2016b, \mn@doi [\mnras] {10.1093/mnras/stw933}, \href
  {http://adsabs.harvard.edu/abs/2016MNRAS.459.4389P} {459, 4389}

\bibitem[\protect\citeauthoryear{{Reeves}, {O'Brien}  \& {Ward}}{{Reeves}
  et~al.}{2003}]{Reeves+03}
{Reeves} J.~N.,  {O'Brien} P.~T.,   {Ward} M.~J.,  2003, \mn@doi [\apjl]
  {10.1086/378218}, \href {http://adsabs.harvard.edu/abs/2003ApJ...593L..65R}
  {593, L65}

\bibitem[\protect\citeauthoryear{{Reeves}, {Braito}, {Nardini}, {Lobban},
  {Matzeu}  \& {Costa}}{{Reeves} et~al.}{2018}]{Reeves+18}
{Reeves} J.~N.,  {Braito} V.,  {Nardini} E.,  {Lobban} A.~P.,  {Matzeu} G.~A.,
   {Costa} M.~T.,  2018, \mn@doi [\apj] {10.3847/2041-8213/aaaae1}, \href
  {https://ui.adsabs.harvard.edu/#abs/2018ApJ...854L...8R} {854, L8}

\bibitem[\protect\citeauthoryear{{Str{\"u}der} et~al.,}{{Str{\"u}der}
  et~al.}{2001}]{Struder+01}
{Str{\"u}der} L.,  et~al., 2001, \mn@doi [\aap] {10.1051/0004-6361:20000066},
  \href {http://adsabs.harvard.edu/abs/2001A%26A...365L..18S} {365, L18}

\bibitem[\protect\citeauthoryear{{Tombesi}, {Cappi}, {Reeves}, {Palumbo},
  {Yaqoob}, {Braito}  \& {Dadina}}{{Tombesi} et~al.}{2010}]{Tombesi+10}
{Tombesi} F.,  {Cappi} M.,  {Reeves} J.~N.,  {Palumbo} G.~G.~C.,  {Yaqoob} T.,
  {Braito} V.,   {Dadina} M.,  2010, \mn@doi [\aap]
  {10.1051/0004-6361/200913440}, \href
  {http://adsabs.harvard.edu/abs/2010A%26A...521A..57T} {521, A57}

\bibitem[\protect\citeauthoryear{{Tombesi}, {Mel{\'e}ndez}, {Veilleux},
  {Reeves}, {Gonz{\'a}lez-Alfonso}  \& {Reynolds}}{{Tombesi}
  et~al.}{2015}]{Tombesi+15}
{Tombesi} F.,  {Mel{\'e}ndez} M.,  {Veilleux} S.,  {Reeves} J.~N.,
  {Gonz{\'a}lez-Alfonso} E.,   {Reynolds} C.~S.,  2015, \mn@doi [\nat]
  {10.1038/nature14261}, \href
  {http://adsabs.harvard.edu/abs/2015Natur.519..436T} {519, 436}

\bibitem[\protect\citeauthoryear{{Zoghbi}, {Fabian}, {Uttley}, {Miniutti},
  {Gallo}, {Reynolds}, {Miller}  \& {Ponti}}{{Zoghbi} et~al.}{2010}]{Zoghbi+10}
{Zoghbi} A.,  {Fabian} A.~C.,  {Uttley} P.,  {Miniutti} G.,  {Gallo} L.~C.,
  {Reynolds} C.~S.,  {Miller} J.~M.,   {Ponti} G.,  2010, \mn@doi [\mnras]
  {10.1111/j.1365-2966.2009.15816.x}, \href
  {http://adsabs.harvard.edu/abs/2010MNRAS.401.2419Z} {401, 2419}

\bibitem[\protect\citeauthoryear{{den Herder} et~al.,}{{den Herder}
  et~al.}{2001}]{denHerder+01}
{den Herder} J.~W.,  et~al., 2001, \mn@doi [\aap] {10.1051/0004-6361:20000058},
  \href {http://adsabs.harvard.edu/abs/2001A%26A...365L...7D} {365, L7}

\makeatother
\end{thebibliography}



%
%
%


\bsp	
\label{lastpage}
\end{document}